\documentclass[12pt]{article}

\usepackage{epsfig}
\usepackage{amssymb}

\newcommand{\app}{\rightarrow}
\newcommand{\ddf}{\delta \Delta F}
\newcommand{\df}{\Delta F}
\newcommand{\dfest}{\Delta F^{\mathrm{est}}}
\newcommand{\dlam}{\Delta \lambda}

\newcommand{\dw}{\Delta W}

\newcommand{\nswitch}{N_{\mathrm{switch}}}
\newcommand{\ntot}{N_{\mathrm{tot}}}

\newcommand{\xbf}{{\mathbf{x}}}

\begin{document}

\title{Extrapolative Analysis of Fast-Switching Free Energy Estimates in a Molecular System}
\author{Daniel M. Zuckerman$^\ast$ and Thomas B. Woolf$^{\ast \dagger}$\\
$^\ast$Department of Physiology and  $^\dagger$Department of Biophysics,\\
Johns Hopkins University School of Medicine, Baltimore, MD 21205\\
\texttt{dmz@groucho.med.jhmi.edu, woolf@groucho.med.jhmi.edu}
}
\date{\today}
\maketitle

\begin{abstract}
We perform an extrapolative analysis of ``fast-growth'' free-energy-difference ($\df$) estimates of a computer-modeled, fully-solvated ethane$\leftrightarrow$methanol transformation.
The results suggest that extrapolation can greatly reduce the systematic error in $\df$ estimated from a small number of very fast switches.
Our extrapolation procedure uses block-averages of finite-data estimates, and appears to be particularly useful for broad, non-Gaussian distributions of data which produce substantial systematic errors with insufficient data.
In every tested case, the extrapolative results were better than direct estimates.
\end{abstract}

\newpage
\section{Introduction}
Relative free energy computations 
have long been of interest, and
biological applications promise to be of particular importance \cite{Beveridge-1989,McCammon-1991,Kollman-1993}.
As examples, it would be desirable to accurately and rapidly estimate free energy changes resulting from the opening of an ion channel, the binding of a ligand, and alchemical mutation among a series of protein ligands.
Ligands might include potential drug compounds or varying sequences of nucleic acids (RNA and DNA).
Strategies for computing free energy differences date back to Kirkwood \cite{Kirkwood-1935} and Zwanzig \cite{Zwanzig-1954} who pioneered thermodynamic integration and free-energy perturbation strategies.
Many computational strategies have since been developed for molecular systems (e.g., \cite{Beveridge-1989,McCammon-1991,Kollman-1993}).

``Fast-growth'' methods \cite{Reinhardt-1992,Reinhardt-1993,Tidor-1997,Jarzynski-1997a,Jarzynski-1997b,Jarzynski-2001,Reinhardt-2000,Hummer-2001} are the focus of the present paper.
The impetus for these approaches comes from the work of Reinhardt, Hunter, and coworkers \cite{Reinhardt-1992,Reinhardt-1993} who recognized that computations could readily employ a microscopic analog of the inequality between work and free energy.
The principle is readily illustrated in an ``alchemical'' context where one wishes to compute the free energy difference between two systems described by different potential energy functions, $U_0$ and $U_1$, and parameterized by the switching variable $\lambda$ according to an extended potential function:
\begin{equation}
\label{ualchem}
U(\xbf;\lambda) = U_0(\xbf) + \lambda [\, U_1(\xbf) - U_0(\xbf) \,]  \; , \hspace{1cm} 0 \leq \lambda \leq 1 \, ,
\end{equation}
where $\xbf$ is a set of configurational coordinates.
If one performs a series of rapid ``switches'' (described below) between the two systems using an amount of work $W$ in each switch, the free energy difference is bounded according to \cite{Wood-1991,Reinhardt-1992}
\begin{equation}
\label{dfbounds}
-\langle W_{1\app 0}\rangle \leq \df_{0\app 1} 
  \leq \langle W_{0\app 1} \rangle \; ,
\end{equation}
where the 
$\langle \cdots \rangle$ brackets 
indicate averages over many switches starting from equilibrium ensembles of either start ($\lambda=0$) or end ($\lambda=1$) systems.
(The distinct ``systems'' could also describe different conformations of a single system constrained to distinct values of a reaction coordinate.)

The potentially rapid, non-equilibrium events used to compute 
$\langle W \rangle$
in Eq.\ (\ref{dfbounds})
thus provide a computational estimate of the equilibrium quantity $\df$.
However, the bounds will not be tight unless the switches are sufficiently slow, offsetting some of the computational savings.

Subsequent work by Jarzynski \cite{Jarzynski-1997a,Jarzynski-1997b} sidesteps, at least in principle, some of the limitations by permitting \emph{direct} computation of $\df$ from a single set of rapid switches, via the simple, \emph{exact} relation, 
\begin{equation}
\label{dfexp}
e^{-\df/ k_B T} 
  = \left\langle e^{-W/ k_B T} \right\rangle \, .
\end{equation}
However, estimates for $\df$ generated using Eq.\ (\ref{dfexp}) are highly sensitive to small values of $W$ and significant errors can arise when the width of the distribution of $W$ values exceeds $k_BT$ \cite{Jarzynski-2001,Hummer-2001}.
Hummer's recent work with a small molecular system concluded that little, if any, advantage was gained from the fast-switching approach \cite{Hummer-2001}.

In the past improvements have been sought in the procedure for generating a set of work values $\{W_1, W_2, \ldots \}$ to be analyzed according to Eq.\ (\ref{dfbounds}) or (\ref{dfexp}).
In particular, one can switch between $\lambda=0$ and 1 along arbitrary paths, perhaps using more than one switching parameter as initially discussed by Reinhardt and coworkers for the fast-switching approach by \cite{Reinhardt-1992,Reinhardt-1993}.
Subsequent exploration of optimal switching paths has been pursued by many workers \cite{Schon-1996,Tidor-1997,Reinhardt-2000,Hummer-2001}.
In fact, the exploration of different paths in alchemical free-energy computations pre-dates the fast-switching approach, and was pursued in free-energy-perturbation and thermodynamic integration efforts --- e.g., \cite{Cross-1986,Berendsen-1986,Kollman-1989}.

\begin{figure}[here]
\begin{center}
\epsfig{file=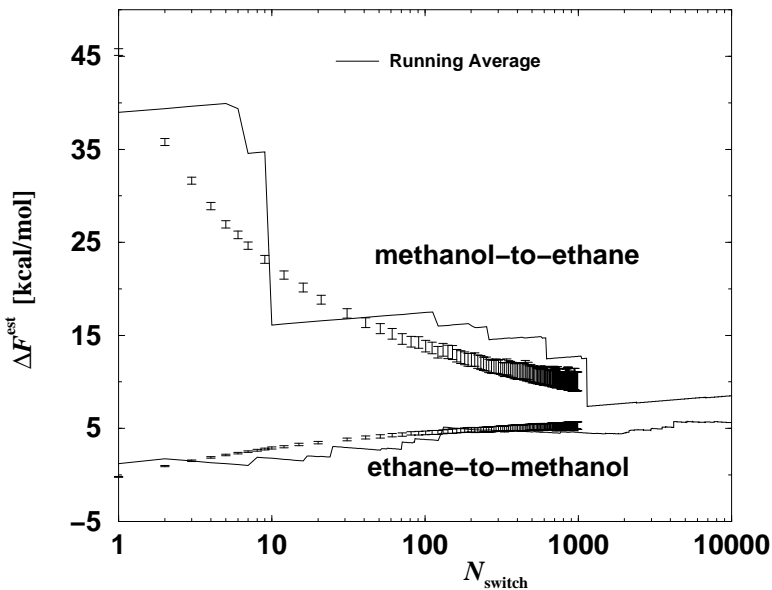,height=4in,angle=270}
\end{center}
\vspace*{-0.5cm}
\caption{\label{fig:run-block}
Running and block averages for forward and reverse switching.
The evolving estimates for the free energy difference are plotted vs.\ the number of switches, $N\equiv\nswitch$.
The running averages [based on Eq.\ (\ref{dfexp}); solid lines] exhibit non-monotonic, rise-and-drop behavior, while the block averages (defined in Sec.\ \ref{sec:block}; error bars) are monotonic and smooth.
Each block-average data point was computed using all 10,000 work ($W$) values.
The error bars are twice the standard error of the mean (see Sec.\ \ref{sec:block}), and represent roughly 90\% confidence intervals \cite{biostats}.
Data from switches of 20 $\lambda$ steps.
}
\end{figure}

The present study, by contrast, attempts to optimize the use of the data $\{W_1, W_2, \ldots \}$ which has already been generated, by using a combination of block-averaging and extrapolation.
This additional statistical analysis is \emph{needed} to bypass the systematic error inherent in finite data samples \cite{Stone-1982,Wood-1991b,Jarzynski-1997b}.
Fig.\ \ref{fig:run-block} illustrates the basic points.
The running averages (solid lines) based on Eq.\ (\ref{dfexp}) exhibit erratic behavior, and it is essentially impossible to judge from these whether the computation has converged to an answer.
However, the same data considered in block-averages (error bars) is well-behaved and, as seen below, well-defined.
Only the block-averages could be considered for extrapolation to the ``infinite-data'' limit.
Cases of insufficient data requiring extrapolation are of great interest because the size of biomolecular systems often makes relative free energy estimates extremely costly.

Following Jorgensen and Ravimohan \cite{Jorgensen-1985} and Jarque and Tidor \cite{Tidor-1997}, we examine alchemical mutations between methanol and ethane in explicit water solvent.
The authors are unaware of any previous application of Jarzynski's relation to \emph{alchemical} transformations in a molecular system, although Hummer performed a methodical study of the inter-methane distance dependence of the free energy \cite{Hummer-2001}.

Our results indicate that the combined use of block-averaging and extrapolation is very promising and warrants additional investigation.
The approach produces successful and reasonably reliable relative-free-energy estimates even from \emph{very fast switches of only one or two steps}, which generate extremely broad, highly-non-Gaussian distributions of work values.
In every case we examined, extrapolation of the data yielded a better estimate than direct averaging alone.

In outline, this Letter is organized as follows:
Sec.\ \ref{sec:genl-alchem} briefly describes ``fast-growth'' computations and gives simulation details.
In Sec.\ \ref{sec:block} we define the block averages, and the extrapolation procedure is discussed in Sec.\ \ref{sec:extrap}.
In Sec.\ \ref{sec:results} we summarize our results and discuss future work, including potential applications of the approach to large biomolecular systems.
We also discuss implications for other approaches to free energy calculations.

\newpage
\section{Alchemical Free Energy Calculations}
\label{sec:genl-alchem}
This section fills in some details regarding the theory governing an ``alchemical'' free energy change and its implementation using a rapid-switching strategy.
Alchemical changes are transformations between Hamiltonians which describe different molecules;
molecular isomerization is mathematically analogous but not considered here.
The free energy difference between the two states is formally given by the ratio of the partition functions according to 
\begin{equation}
\label{dfratio}
\exp{\left( -\df_{0\app 1} / k_B T \right)}
	= \frac{ \int d \xbf \, e^{-U_1(\xbf)/ k_B T}  }
	{ \int d \xbf \, e^{-U_0(\xbf)/ k_B T}  } \, .
\end{equation} 
Jarzynski's relation (\ref{dfexp}) is derived from this definition.

\subsection*{Free Energy Perturbation (FEP)}
The so-called free-energy-perturbation (FEP) procedure for computing relative free energies \cite{Kirkwood-1935,Zwanzig-1954,Valleau-1972} is a well-established method for molecular systems \cite{Berendsen-1986,Kollman-1989,Beveridge-1989,McCammon-1991,Kollman-1993} which we use as a benchmark for understanding systematic errors.
FEP computations entail a number of equilibrium simulations performed at a set of \emph{fixed} values of $\lambda$;
for example, our FEP result quoted in Sec.\ \ref{sec:results} uses simulations at $\lambda = 0.1, 0.2, \ldots, 0.9$.
The total free energy change is estimated as the sum of the incremental changes, which are computed based on the analog of Eq.\ (\ref{dfexp}) involving 
$\langle \exp{(-\dw / k_B T)} \rangle$, where $\dw$ is the work or energy difference between configurations at different $\lambda$ values.

\subsection*{Fast-Growth Procedure}
Fast-growth algorithms have been discussed in detail elsewhere (e.g., \cite{Jarzynski-1997a,Jarzynski-1997b,Jarzynski-2001,Hummer-2001}), so we merely sketch the approach.
The general procedure for computing a ``fast-growth'' free energy difference --- via Eq.\ (\ref{dfexp}) rather than (\ref{dfratio}) --- begins with the generation of an equilibrium ensemble of starting ($\lambda=0$) configurations, perhaps by molecular dynamics simulation as is done here.
One proceeds by (i) choosing a configuration from the equilibrium ensemble, (ii) incrementing the potential energy function (\ref{ualchem}) to a new, greater value of $\lambda$ (keeping the configuration fixed) and (iii) relaxing the system at the new $\lambda$ value.
Steps (ii) and (iii) are repeated until a value $\lambda \lesssim 1$ is reached.
In our implementation, the $\lambda$ increments in (ii) are uniform and the relaxation stage (iii) consists of a single molecular dynamics (MD) ``relaxation'' step, following the ``fast-growth'' convention \cite{Reinhardt-1992,Jarzynski-2001}.
A uniform increment of $\dlam = 0.05$, for instance, corresponds to 20 ``$\lambda$ steps'' and would require 19 MD steps, as none is necessary at $\lambda=1$.

The work for any such switch is computed based only on the potential energy increments and not the relaxation dynamics.
Thus, if $\xbf_i^{\mathrm{fin}}$ denotes the final configuration of the system after it is relaxed at the $i$th value $\lambda_i$, the work calculated from
\begin{equation}
\label{work}
W = \sum_{i=1} \left[ U \! \left(\xbf_{i-1}^{\mathrm{fin}};\lambda_i\right) - U \! \left(\xbf_{i-1}^{\mathrm{fin}};\lambda_{i-1}\right) \right ] \;,
\end{equation}
where \emph{the same configuration} is evaluated at two different $\lambda$ values.
Finally, to evaluate the averages in Eqs.\ (\ref{dfbounds}) and (\ref{dfexp}), one uses additional members of the $\lambda=0$ equilibrium ensemble to generate subsequent values of $W$ --- starting from step (i), above.

\subsection*{Methanol$\leftrightarrow$Ethane Model and Simulation}
Simulations of the methanol$\leftrightarrow$ethane ``transmutation'' were performed within the CHARMM molecular dynamics package.
Both methanol and ethane were modeled in the united-atom picture: methanol was represented as a three-atom (C,O,H) molecule and ethane as a two-atom (C,C) molecule.
The solvent used 125 TIP3 water molecules (for both $\lambda=0$ and 1) in a periodically replicated box of (15.6 \AA)$^3$.  
To facilitate comparison with earlier studies, electrostatics and van der Waals interactions were both shifted to zero at a cutoff of 8 \AA.  
Molecular dynamics steps (performed at fixed $\lambda$ values) used the leapfrog Verlet algorithm.
The same simulation procedure and parameters were used for free energy perturbation calculations.

\newpage
\section{Block Averaging}
\label{sec:block}
While block-averaging is straightforward, its \emph{repeated application for growing block sizes to a non-linear transformation} --- such as taking the log of an average of exponentials in Jarzynski's relation (\ref{dfexp}) --- turns out to yield rich, well-behaved data: see Fig.\ \ref{fig:run-block}.
The procedure and some implications are discussed now.

We construct block averages \cite{Wood-1991b,bootstrap,Jarzynski-1997b,subsampling} from a set of, say, $\ntot$ work values \\ 
$\{W_1, W_2, \ldots, W_{\ntot}\}$
by applying Jarzynski's relation (\ref{dfexp}) to a series of blocks, each containing $N \equiv \nswitch$ values.
More specifically, we define the $N$-block-averaged estimate for the free energy as
\begin{equation}
\label{dfblock}
\df_N = \frac{N}{\ntot} \sum_{n=1}^{\ntot/N} -k_BT \log{\langle e^{-W/k_BT} \rangle_{N,n} } 
\equiv \langle f_N \rangle \,,
\end{equation}
where the individual block averages are defined by
\begin{equation}
\label{wblock}
\langle e^{-W/k_BT} \rangle_{N,n} 
  = \frac{1}{N} \sum_{i=(n-1)N+1}^{nN} e^{-W_i/k_BT} 
\equiv f_{N,n} \, .
\end{equation}
The ratio $\ntot/N$ denotes the largest integer less than or equal to the literal fraction, and is never less than 30 in our analysis.
Because of potential correlations in the sequence
$\{W_1, W_2, \ldots, W_{\ntot}\}$
we randomly re-sort the values prior to computing the block results presented here.
We note that larger block sizes, $N$, could be considered with a bootstrap \cite{bootstrap} or subsampling \cite{subsampling} analysis.

The true free energy difference of Eq.\ (\ref{dfexp}) is $\df = \df_\infty$, and the other limit gives the average work, $\langle W \rangle = \df_1$: see Fig.\ \ref{fig:run-block}.
In general, a finite value of $N$ indicates that the average in Eq.\ (\ref{wblock}) is performed from a poor sample of the $W$ distribution, with $N$ determining how much of the tails of the distribution are included in the average.
However, the averaging of these poor samples in Eq.\ (\ref{dfblock}) yields a well-defined descriptor of the finite-$N$ statistics.
In the present case, the Boltzmann-factor form ensures monotonic behavior,
so that 
\begin{equation}
\label{dfn_ineq}
\df_{N+1} \leq \df_N \,, 
\end{equation}
the essence of which was noted by Jarzynski \cite{Jarzynski-1997b}; 
see also \cite{Wood-1991b}.
The usual relation between the average work and free energy (\ref{dfbounds}) is simply a weaker case of the more general inequality (\ref{dfn_ineq}).

The uncertainty in the finite-$N$ free energy values, $\ddf$, is estimated by twice the standard error of the mean,
\begin{equation}
\label{conf}
(\ddf_N)^2 = \frac{4}{(\ntot/N)^2} \sum_{n=1}^{\ntot/N} 
  \left( f_{N,n} - \langle f_N \rangle \right)^2 \, ,
\end{equation}
which gives roughly a 90\% confidence interval \cite{biostats}.
This is the quantity used to compute error bars and uncertainties.

\newpage
\section{Extrapolation}
\label{sec:extrap}
While extrapolation and data-fitting are something of black arts, one can hope to derive meaningful information with a careful error analysis \cite{numerical-recipes}.
Here we discuss some simple, intuitively appealing schemes for extrapolating the block-averaged, finite-data free energies (\ref{dfblock}) to the limit of infinite data.
The motivation for our approach is the analysis of finite-\emph{size} effects in spin systems \cite{Fisher-1971,Binder-1997}.

Inspection of the data on a linear scale, such as Fig.\ \ref{fig:run-block}, and in logarithmic plots suggests the simplest fit might be to a power law, 
\begin{equation}
\label{power-fit}
\df_N = \df_\infty + a_1 (1/N)^{\alpha_1} \, .
\end{equation}
A natural, related form considers a power series
\begin{equation}
\label{series-fit}
\df_N = \df_\infty + \sum_{k=1}^{k_{max}} b_k (1/N)^{k \beta_1} \, ,
\end{equation}
where the parameter $\beta_1$ can be chosen from a fit or some other way, such as by examining the leading $1/N$ behavior.
Our work with the form (\ref{series-fit}) uses three parameters with $k_{max} = 2$, except where noted, and the fixed exponent $\beta_1 = 0.266$ chosen empirically, but based on some of the values fitted for $\alpha_1$ in Eq.\ (\ref{power-fit}).
Naturally other exponents and polynomial degrees could be used.

One drawback to these forms is clear: 
if the data do not include the leading $1/N$ behavior and the ``distance'' to extrapolate is great (from $1/N=0$ to the first data point; see Fig.\ \ref{fig:extrap}), the fits will not have good extrapolative power.
We anticipate that an analytic understanding of the behavior of $\df_N(N)$ for model systems, to be pursued in future work, will shed light on extrapolation forms and methods.

\newpage
\section{Results}
\label{sec:results}
We now present estimates for the free energy difference of the methanol$\rightarrow$ethane transformation, based on the block-averaging and extrapolation presented in the previous two sections.
Our focus is the methanol-to-ethane direction of the transformation because it is more challenging and so presumably a better model for larger systems.

The basic results are surprising and exciting.
First, successful extrapolation to reasonably accurate free energy values does appear to be possible in the methanol$\rightarrow$ethane system.
Moreover, for fixed amounts of computer time, the extrapolated estimates appear to be considerably better than standard fast-growth values, and can avoid errors of several kcal/mole resulting from insufficient data.
If borne out for other systems, the ability to make estimates from a relatively small number of very rapid switches would mean dramatic efficiency gains.

\begin{figure}[here]
\begin{center}
\epsfig{file=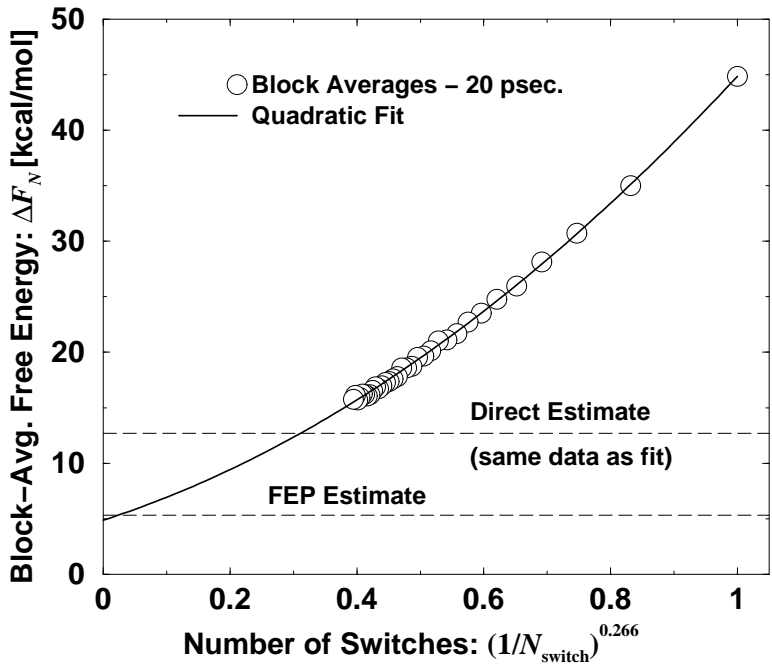,height=4in,angle=270}
\end{center}
\vspace*{-0.5cm}
\caption{\label{fig:extrap}
\small 
Extrapolation of the free energy estimate for a fully-solvated methanol$\rightarrow$ethane transformation. 
Finite-switch averages $\df_N$ are plotted as a function of the number of switches per average $N\equiv\nswitch$ raised to a ``scaling'' power.
The block averages, fit, and direct estimate were all computed from the same data, while the free-energy-perturbation (FEP) value was generated from an independent, substantially longer calculation.
The symbols roughly indicate the sizes of the error bars.
The block-averaging is described in Sec.\ \ref{sec:block}, the extrapolative fitting in Sec.\ \ref{sec:extrap}, and the direct estimate averages the same data according to Eq.\ (\ref{dfexp}).
The data are from $10^3$ switches of 20 $\lambda$ steps each, and note that $k_B T \simeq 0.6$ kcal/mole.
}
\end{figure}
Fig.\ \ref{fig:extrap} shows a sample extrapolation, based on Eq.\ (\ref{series-fit}), for $10^3$ switches of 20 $\lambda$ steps each.
Note that the \emph{un-}extrapolated, ``direct'' free-energy estimate --- based on application of Eq.\ (\ref{dfexp}) to the same data --- exceeds both the extrapolated value and the reliable FEP estimate by 7 kcal/mole$ \simeq 11 k_BT$. 
Thus, with a limited amount of data, extrapolation of the block-averaged values yields a much better estimate.

\begin{table}
\caption{\label{table} Extrapolated estimates for the solvated methanol$\rightarrow$ethane free energy difference.
The estimates $\dfest$ and uncertainties are given in units of kcal/mole, and may be compared to the free-energy-perturbation estimate of 5.3 $\pm 0.16$ kcal/mole.
Direct estimates are computed from Eq.\ (\ref{dfexp})
and power-series extrapolations from (\ref{series-fit}), using identical data.
The uncertainties in the extrapolations are discussed in Sec.\ \ref{sec:results}.
The bracketed values give the differences between the direct $\df$ estimates and the more costly FEP estimate in the first row, and hence measure the accuracy of the former.
The quantity ``$\lambda$ steps'' indicates the number of increments in the alchemical coordinate: see Sec.\ \ref{sec:genl-alchem}.
``Total Steps'' gives the number of MD steps excluding those for generating the equilibrium ensemble at $\lambda=0$.
}
\begin{center}
\begin{tabular}{lcccc}
\hline \hline
Method 			& $\dfest$ 	& Uncert'y & $\lambda$ steps
							& Tot.\ Steps\\
\hline
Direct			&	7.37	&[2.1]	& 200	& $2\cdot10^6$ \\
Extrapolation		&	5.68	& 1.66	& 200	& $2\cdot10^6$ \\
\\
Direct			&	11.2	&[5.9]	& 200	& $2\cdot10^5$ \\
Direct			&	8.50	&[3.2]	& 20	& $2\cdot10^5$ \\
Extrapolation		&	4.93	& 0.960	& 200	& $2\cdot10^5$ \\
Extrapolation		&	6.41	& 1.21	& 20	& $2\cdot10^5$ \\
\\
Direct			&	12.7	&[7.4]	& 20	& $2\cdot10^4$ \\
Direct			&	8.68	&[3.4]	& 2	& $1\cdot10^4$ \\
Extrapolation		&	4.87	& 1.08	& 20	& $2\cdot10^4$ \\
Extrapolation		&	$\;7.85^*$	
					& $\;1.46^*$	
						& 2	& $1\cdot10^4$ \\
\hline \hline
\end{tabular}
\end{center}
$^*$ These values change to 5.03 and 1.85 for $k_{max}=3$ in Eq.\ (\ref{series-fit}) with a substantially improved goodness-of-fit measure.
\end{table}

Table \ref{table} presents quantitative results for the methanol-to-ethane transformation.
The extrapolations are uniformly superior to the direct estimates for any fixed amount of computer time and consistently avoid errors on the order of several kcal/mole (where 1 kcal/mole $\simeq 1.6 \, k_BT$) for smaller amounts of data.
Total computer times for the tabulated results range from 2 nsec.\ ($10^4$ switches of 200 steps) down to just 10 psec ($10^4$ 2-step switches) of non-equilibrium molecular dynamics simulation.
The ``Total Steps'' column does not include the computer time expended on generating an equilibrium ensemble at $\lambda=0$ because it is unlikely that one would investigate the transmutation of a system which has not already been subjected to an equilibrium study.

We estimated upper and lower bounds simply by extrapolating, independently, from the sets of upper and lower limits of the confidence intervals; recall Eq.\ (\ref{conf}).  
Statistical uncertainties were not given for the direct estimates because the systematic error is clearly more significant than the statistical:
the bracketed deviations in Table \ref{table} indicate the direct estimates differ dramatically from the free-energy perturbation (FEP) result.
Recall that the FEP approach was outlined in Sec.\ \ref{sec:genl-alchem}.

The power of the extrapolative approach is underscored by the challenging character of the distributions of work values under consideration.
The distributions are all quite broad and asymmetric:
standard deviations range from 12 kcal/mole $\simeq 20 \, k_BT$ (for the 200-step switches) to 24 kcal/mole $\simeq 38 \, k_BT$ (2 steps), and third moments range from 72\% of the standard deviation (200 steps) to 100\% (2 steps).
Thus, although all of the tabulated simulations involve very rapid switches --- of less than 1 psec.\ of molecular dynamics time per switch --- the substantial differences in the distributions indicate that the data sets are quite distinct.
We also noted a degree of robustness in trials with related but different forms and exponents $\beta_1$ (results not shown) which typically yielded consistent, if slightly inaccurate, results across data sets from widely disparate numbers of $\lambda$ steps --- and hence disparate computer times and work distributions.

Despite the success of the fitting form used here, superior extrapolations may be possible.
The forms employed here (see Sec.\ \ref{sec:extrap}) are empirical, so a theoretical basis should provide additional insight.
Lacking that, a more systematic exploration of the implicit parameters --- the exponent $\beta_1$ in (\ref{series-fit}), the minimum number of switches per block, and the degree of the fitting polynomial --- would also be valuable.

Another interesting trend illustrated in the data of Table \ref{table} is that for a fixed amount of computer time, \emph{direct} estimates using fewer $\lambda$ steps appear to give better results.
The statistical errors (data not shown) are also better for direct estimates using more rapid switches.

\newpage
\section{Summary and Discussion}
We have performed and analyzed extrapolative free energy estimates based on ``fast-growth'' alchemical simulations of a fully solvated methane$\leftrightarrow$ethanol transformation.
The results of Table \ref{table} suggest that the combined use of block-averaging (Sec.\ \ref{sec:block}) and extrapolation (Sec.\ \ref{sec:extrap}) permits accurate estimates from a relatively small amount of data which --- when analyzed using the standard ``direct'' method --- leads to unacceptably large systematic errors of several kcal/mole.
Extrapolated results, for our system, were \emph{always} better than standard, direct estimates.
The approach also appears to be fairly robust, in that good results are achieved over ranges both of overall computer time and of alchemical switching speeds.

Our work builds on that of Wood \emph{et al.}, who perceptively proposed a first-order estimate of the systematic errors due to finite samples of data \cite{Wood-1991b}.
The present method, however, is not limited to narrow work (energy-change) distributions as noted in Sec.\ \ref{sec:results}.

This Letter describes an initial exploration of a potentially important approach, and a number of important issues and questions deserve further exploration.
To name a few:
(i) undoubtedly, simultaneous fits of forward ($\lambda = 0 \! \rightarrow \! 1$) and reverse switching data will provide more reliable free energy estimates; 
(ii) we have not performed a quantitative analysis of the efficiency, both by comparison to standard ``fast-growth'' approaches as well as to free-energy-perturbation estimates;
(iii) how does the extrapolation approach generalize to larger biomolecular systems?
(iv) how universal are the behaviors of the finite-data estimates, $\df_N$, considered in the extrapolation?
(v) can theoretical scrutiny of simple models and distributions clarify the extrapolative procedure?
The ideas discussed here may also apply, with suitable modifications, to perturbative calculations.

We have discussed methods for analyzing data from fast-switching simulations, but have not broached the possibilities for improved \emph{sampling} of data.
There appear to be a number of promising, unexplored avenues.
Instead of using a uniform alchemical increment $\dlam$, for example, one could adjust increments to ensure relatively constant work increments, following the example of perturbative calculations \cite{Cross-1986,Berendsen-1986,Kollman-1989,Grossfield-2001};
this approach could also be adapted for higher-dimensional alchemical coordinates already considered by others \cite{Reinhardt-1992,Reinhardt-1993,Brooks-1996,Tidor-1997}.
Improved sampling efficiency may also result from biasing the ``relaxational,'' fixed-$\lambda$ dynamics to favor states with smaller work increments.

Finally, we note that the relationship between the approach described here and established statistical methods needs to be elucidated.
Elements of our approach, particularly the construction of ``finite-data'' block averages, clearly have been considered in ``bootstrap'' \cite{bootstrap} and ``subsampling'' \cite{subsampling} statistical approaches.
Nevertheless, the authors are not aware of a similar practical --- if \emph{ad hoc} --- technique for extrapolation to the infinite-data limit like that presented here.

\section*{Acknowledgments}
Many people provided helpful comments and suggestions for the research reported here.
The authors would like to thank Lucy Forrest, Lancelot James, Chris Jarzynski, Hirsh Nanda, Horia Petrache, Lawrence Pratt, Mark Robbins, Jonathan Sachs, Thomas Simonson, Scott Zeger, and David Zuckerman.
Funding for this work was provided by the NIH (Grant GM54782),
the Bard Foundation, and the Department of Physiology.
D.M.Z. is the recipient of a National Research Service Award (GM20394).

\newpage

\begin{thebibliography}{10}

\bibitem{Beveridge-1989}
D.~Beveridge and F.~Di{C}apua.
\newblock Free energy via molecular simulation: applications to chemical and
  biomolecular systems.
\newblock {\em Ann. Rev. Biophys. Biophys. Chem.}, 18:431--492, 1989.

\bibitem{McCammon-1991}
J.~A. McCammon.
\newblock Free energy from simulations.
\newblock {\em Curr Opin. Struc. Bio.}, 2:96--200, 1991.

\bibitem{Kollman-1993}
P.~A. Kollman.
\newblock Free energy calculations: Applications to chemical and biochemical
  phenomena.
\newblock {\em Chemical Reviews}, 93:2395--2416, 1993.

\bibitem{Kirkwood-1935}
J.~G. Kirkwood.
\newblock Statistical mechanics of fluid mixtures.
\newblock {\em J. Chem. Phys.}, 3:300--313, 1935.

\bibitem{Zwanzig-1954}
R.~W. Zwanzig.
\newblock High-temperature equation of state by a perturbation method.
\newblock {\em J. Chem. Phys.}, 22:1420--1426, 1954.

\bibitem{Reinhardt-1992}
W.~P. Reinhardt and J.~E. Hunter.
\newblock Variational path optimization and upper and lower bounds to free
  energy changes via finite time minimization of external work.
\newblock {\em J. Chem. Phys.}, 97:1599--1601, 1992.

\bibitem{Reinhardt-1993}
J.~E. Hunter, W.~P. Reinhardt, and T.~F. Davis.
\newblock A finite-time variational method for determining optimal paths and
  obtaining bounds on free energy changes from computer simulations.
\newblock {\em J. Chem. Phys.}, 99:6856--6864, 1993.

\bibitem{Tidor-1997}
C.~Jarque and B.~Tidor.
\newblock Computing bounds on free energy changes with one and two dimensional
  paths.
\newblock {\em J. Phys. Chem. B}, 101:9402--9409, 1997.

\bibitem{Jarzynski-1997a}
C.~Jarzynski.
\newblock Nonequilibrium equality for free energy differences.
\newblock {\em Phys. Rev. Lett.}, 78:2690--2693, 1997.

\bibitem{Jarzynski-1997b}
C.~Jarzynski.
\newblock Equilibrium free-energy differences from nonequilibrium measurements:
  A master equation approach.
\newblock {\em Phys. Rev. E}, 56:5018--5035, 1997.

\bibitem{Jarzynski-2001}
D.~A. Hendrix and C.~Jarzynski.
\newblock A {``}fast growth{''} method of computing free energy differences.
\newblock {\em J. Chem. Phys.}, 114:5974--5981, 2001.

\bibitem{Reinhardt-2000}
M.~A. Miller and W.~P. Reinhardt.
\newblock Efficient free energy calculations by variationally optimized metric
  scaling: Concepts and applications to the volume dependence of cluster free
  energies and to solid-solid phase transitions.
\newblock {\em J. Chem. Phys.}, 113:7035--7046, 2000.

\bibitem{Hummer-2001}
G.~Hummer.
\newblock Fast-growth thermodynamic integration: Error and efficiency analyis.
\newblock {\em J. Chem. Phys.}, 114:7330--7337, 2001.

\bibitem{Wood-1991}
R.~H. Wood.
\newblock Estimation of errors in free energy calculationd due to the lag
  between the hamiltonian and system configuration.
\newblock {\em J. Phys. Chem.}, 95:4838--4842, 1991.

\bibitem{Schon-1996}
J.~C. Sch{\"{o}}n.
\newblock A thermodynamic distance criterion of optimality for the calculation
  of free energy changes from computer simulations.
\newblock {\em J. Chem. Phys.}, 105:10072--10083, 1996.

\bibitem{Cross-1986}
A.~J. Cross.
\newblock A comment on hamiltonian parameterization in kirkwood free energy
  calculations.
\newblock {\em Ann. N.Y. Acad. Sci}, 482:89--90, 1986.

\bibitem{Berendsen-1986}
T.~P. Straatsma, H.~J.~C. Berendsen, and J.~P.~M. Postma.
\newblock Free energy of hydrophobic hydration: A molecular dynamics study of
  noble gases in water.
\newblock {\em J. Chem. Phys.}, 85:6720--6727, 1986.

\bibitem{Kollman-1989}
D.~A. Pearlman and P.~A. Kollman.
\newblock A new method for carrying out free energy perturbation calculations:
  Dynamically modified windows.
\newblock {\em J. Chem. Phys.}, 90:2460--2470, 1989.

\bibitem{Stone-1982}
A.~D. Stone and J.~D. Joannopoulos.
\newblock Finite ensemble averages of the zero-temperature resistance and
  conductance of disordered one-dimensional systems.
\newblock {\em Phys. Rev. E}, 25:2400--2404, 1982.

\bibitem{Wood-1991b}
R.~H. Wood, W.~C.~F. M{\"{u}}hlbauer, and P.~T. Thompson.
\newblock Systematic errors in free energy perturbation calculations due to a
  finite sample of configuration space: Sample-size hysteresis.
\newblock {\em J. Phys. Chem.}, 95:6670--6675, 1991.

\bibitem{biostats}
Daniel~W. W.
\newblock {\em Biostatistics}.
\newblock Wiley, {New York}, 1974.

\bibitem{Jorgensen-1985}
W.~L. Jorgensen and C.~Ravimohan.
\newblock Monte {C}arlo simulation of differences in free energies of
  hydration.
\newblock {\em J. Chem. Phys.}, 83:3050--3054, 1985.

\bibitem{Valleau-1972}
J.~P. Valleau and D.~N. Card.
\newblock Monte {C}arlo estimation of the free energy by multistage sampling.
\newblock {\em J. Chem. Phys.}, 57:5457--5462, 1972.

\bibitem{bootstrap}
B.~Efron and R.~J. Tibshirani.
\newblock {\em An Introduction to the Bootstrap}.
\newblock Chapman and Hall, {New York}, 1993.

\bibitem{subsampling}
D.~N. Politis, J.~P. Romano, and M.~Wolf.
\newblock {\em Subsampling}.
\newblock Springer, {New York}, 1999.

\bibitem{numerical-recipes}
W.~H. Press, S.~A. Teukolsky, W.~T. Vetterling, and B.~P. Flannery.
\newblock {\em Numerical Recipes in Fortran}.
\newblock Cambridge University Press, Cambridge, England, 2nd edition, 1992.

\bibitem{Fisher-1971}
M.~E. Fisher.
\newblock The theory of critical point singularities.
\newblock In M.~S. Green, editor, {\em {Critical Phenomena}}, pages 1--99, {New
  York}, 1971. Academic.

\bibitem{Binder-1997}
K.~Binder and D.~W. Heermann.
\newblock {\em {Monte Carlo Simulation in Statistical Physics}}.
\newblock Springer, Berlin, 1997.

\bibitem{Grossfield-2001}
A.~Grossfield and T.B. Woolf.
\newblock {Interaction of Tryptophan Analogs with POPC Lipid Bilayers
  Investigated by Molecular Dynamics Calculations}.
\newblock Submitted for publication.

\bibitem{Brooks-1996}
X.~Kong and C.~M. Brooks.
\newblock Lambda-dynamics: A new approach to free energy calculations.
\newblock {\em J. Chem. Phys.}, 105:2414--2423, 1996.

\end{thebibliography}

\end{document}